# A Greedy Double Swap Heuristic for Nurse Scheduling

Murphy Choy[1] and Michelle Cheong

*Singapore Management University, School of Information System*

*80 Stamford Road, Singapore 178902*

Email: murphychoy@smu.edu.sg; michcheong@smu.edu.sg

## Abstract

One of the key challenges of nurse scheduling problem (NSP) is the number of constraints placed on preparing the timetable, both from the regulatory requirements as well as the patients' demand for the appropriate nursing care specialists. In addition, the preferences of the nursing staffs related to their work schedules add another dimension of complexity. Most solutions proposed for solving nurse scheduling involve the use of mathematical programming and generally considers only the hard constraints. However, the psychological needs of the nurses are ignored and this resulted in subsequent interventions by the nursing staffs to remedy any deficiency and often results in last minute changes to the schedule. In this paper, we present a staff preference optimization framework which is solved with a greedy double swap heuristic. The heuristic yields good performance in speed at solving the problem. The heuristic is simple and we will demonstrate its performance by implementing it on open source spreadsheet software.

## 1.0   Introduction

Nurse scheduling is a critical issue that many hospitals face in their operations. The main objective of nurse scheduling is to prepare a work schedule that meets manpower needs of the various wards while satisfying the various legislation requirements, as well as workload distribution among the nurses. Very often, the nurses will indicate their individual preferences to the nurse manager who will try to incorporate these preferences within the schedule. Occasionally, some private arrangements were done through nurses' social networks and informal discussions which further encumber the arrangement process. This makes the entire process iterative and very time consuming.

The nurse scheduling problem is a well explored NP-hard combinatorial optimization problem in literature and there are many researchers who have explored many ways to highlight the plausible solution space (Osogami and Imai, 2000; Brucker et. al., 2010). The early research focused on a scaled down version of the problem that can be solved using mathematical programming techniques (Miller et. al., 1976). These scaled down formulation of the problem are too small and often ignored too many constraints to be of any practical value. Given the limitations of mathematical programming techniques, many research works branched out into other optimization approaches such as constraint programming (Okada & Okada, 1988), expert system (Chen and Yeung, 1993) and heuristics (Dowsland, 1998; Brusco and Jacobs, 1993; Easton and Mansour, 1993).

---

[1] Corresponding Author.

Existing literature constructed the objective function based on the information from other literature as well as consultation with the planning staffs. This resulted in the assumption that the objectives of the models have taken care of the preferences of the staffs and thus do not provide for flexibility in the re-arrangement of the schedules once they have been created. There are views that the assumption of the models incorporating most of the preferences as a strong one, however, in general, it is not the case. Most importantly, none of the existing literature mentioned about the problem of ad-hoc changes that can occur that necessitate changes to the schedule.

The problem in this paper is highlighted by one of the public hospital in Singapore who is having difficulties in planning their nursing schedule. For this problem, we are examining a schedule involving 15 nurses over a 2-week period, for the intensive care unit. The key purpose of this paper is to solve a realistic NSP problem while managing the preferences of the nurses. We are proposing a greedy double swap heuristic in which a first greedy swapping algorithm is employed to obtain an initial solution, followed by a second greedy swapping algorithm to optimize the preferences of the nurses. The first heuristic will optimize by satisfying the hard constraints to reach a sub-space of possible solutions. Once the heuristic finds the sub-space, it then initiates the second heuristic to search within the sub-space for solutions to satisfy the soft constraints which minimizes the penalty cost objective function. The proposed heuristic allows for a closer approximation to the preferences of the nurses by accepting inputs from them. The nurses inputs are essentially cost assigned to different soft constraints, where a higher cost imply lower preference, while a lower cost imply higher preference. The heuristics can be implemented easily using open source spreadsheet tools which support spreadsheet programming.

The paper is structured as follows. Section 2 provides a short introduction to the NSP problem and the prior research done on the problem. Section 3 will focus on the formulation of the NSP problem. Section 4 describes the Greedy Double Swap Heuristic and its pseudo-code. Section 5 presents the experimental results on both a mock up scenario as well as the original problem using an open source spreadsheet tool. Finally, in section 6, we will conclude the results of this study.

## 2.0 Background

Many real world problems arise due to constraints on the resource and demand. Therefore, proper management of the constraints has been a keystone issue in optimization research. There are attempts to manage these through the use of constraints programming, heuristics as well as mathematical programming. Interested readers can refer to these papers Coello-Coello (2002) and Osogami and Imai (2000).

Most NSPs are complex with a high number of hard and soft constraints. There are discussions that the problem is more complex than other optimization problems such as the traveling salesman problem (Tien and Kamiyama, 1982). The NSPs problems are typically decomposed into three stages, Morning, Afternoon and Night shifts. There are other forms of decomposition such as the one proposed in Dowsland and Thompson (2000). The Knapsack approach initially checks whether the supply of nurses matches the demand requirements. If the demand exceeds the supply, the focus will be on the manpower requirements and meeting as much of the requirements as possible. Once the manpower requirements are met or should the supply exceeds the demand, the next stage will be on the appropriate assignment of the various day, afternoon and night shifts. The most common heuristics used in these assignments is Tabu-Search. Experiments have shown this approach to be quite effective and other heuristics can be applied in stage 2 optimization of hard and soft constraints as well.

In the case of genetic algorithms, special encoding has to be employed in order to make the problem solvable. The solution space is encoded as chromosomes using the decoder-index string form which can be solved using the indirect genetic algorithms. Once solved, each chromosome is then applied to the nurse's schedule (Aickelin and Dowsland, 2004). Similar encodings are adopted in rule-based approaches which attempt to mimic a human thinking process in solving NSP problems (Li and Aickelin, 2003). The approach was compared against human schedulers' solution and there were discussions about the ability of the human scheduler being able to identify the most appropriate solution or approach given the existing gap in the schedule. To address this problem, Bayesian Optimization approach was proposed to reduce the gap by acting as the human mediation factor. Both approaches are able to produce feasible solutions faster than other comparable methods.

Recent research have focused on variants of the genetic algorithms as well as improved mathematical programming techniques. New techniques such as Interactive Genetic Algorithms (You, Yu and Lien, 2010), Component-Based Heuristic Search Method (Li, Aickelin and Burket, 2008) and Mixed Integer Non-Linear Programming approach (Azaiez and Sharif, 2005).

## 3.0   Problem Definition

One of the main challenges in planning is the scheduling of nurses. Different types of nurses are needed for different types of ward. One of the nurse scheduling problems faced by the hospital is to create a bi-weekly schedule for the intensive care wards which are taken care by 15 nurses with various specializations. The schedule produced must satisfy the hard constraints such as obeying the regulations, be contractually legal, meeting the requirements of the special wards, satisfying the demand for the nurse in the ward, and ensuring that the leave plans of the nurses are met. Other soft constraints such as not to assign shifts which are too close together (e.g. afternoon shift followed by morning shift the next day), male nurses cannot be assigned to female ward, language ability of the nurses to take care of patients of different ethnicity, are also to be adhered to, if possible.

A three shifts system is implemented here where a single day is split into morning, afternoon and night shifts. For scheduling purpose, morning shift is represented by 1, afternoon shift by 2 and night shift by 3. Rest days are marked as 0. Generally, the nurses do not work the same shift throughout the 14 days. The nurses will rotate the shifts to ensure even distribution of work duties among all the nurses. Unless specifically requested, the planning staffs will try to ensure that everyone has the same number of different shifts. There are 5 grades of nurses, namely Assistant Nurse (AN), Staff Nurse 1 (SN1), Staff Nurse 2 (SN2), Senior Staff Nurse 1 (SSN1) and Senior Staff Nurse 2 (SSN2). SSN2 is the most senior while AN is the most junior.

While in other literature, there are discussions about the ability of the senior nurses covering the junior nurses (Dowsland, 1998; You, Yu and Lien, 2010), this discussion is incompatible due to the special duty requirements for senior nurses which are regulated by legislation and that the duties of a junior nurse cannot be covered by a senior nurse.

In our problem, the hard constraints which control the feasibility of the solutions are given as follow.

**H1**: Minimum Rest Days Constraint

The number of rest days for a schedule of D days, must be at least $G_{min}$ days

$$\sum_j \sum_s (w_{i,s,j} = 0) \geq G_{min}, \forall\, i = \{1, \ldots, n\} \quad (1)$$

> Where,
> $G_{min}$ = the minimum number of days off in a period of D
> $w_{i,s,j}$ = 1 if nurse i is working on shift s on day j, and 0 otherwise

**H2**: Maximum Consecutive Work Days with Rest Day constraint

For every K consecutive work days, there must be at least 1 rest day.

$$\prod_{j}^{j+K-1} w_{i,s,j} = 0,\ \forall\, i = \{1, \ldots, n\},\ \forall\, s = \{1,2,3\},\ \forall\, j = \{1, \ldots, D-K+1\} \quad (2)$$

> Where,
> K = maximum number of consecutive working days for a schedule of D days
> $w_{i,s,j}$ = 1 if nurse i is working on shift s on day j, and 0 otherwise

This constraint is modified if the consecutive work days are all night shifts. After 3 consecutive night shifts, there must be 1 sleep day and 1 rest day (equivalent to 2 rest days, but only the 2$^{nd}$ rest day will be counted towards $G_{min}$ as defined in constraint 1).

$$\prod_{j}^{j+3} w_{i,3,j} = 0, \ \forall\, i = \{1, \ldots, n\},\ \forall\, j = \{1, \ldots, D - 4 + 1\} \quad (2a)$$

$$\prod_{j}^{j+4} w_{i,3,j} = 0, \ \forall\, i = \{1, \ldots, n\},\ \forall\, j = \{1, \ldots, D - 4 + 1\} \quad (2b)$$

And, after 4 consecutive night shifts, there must be 1 sleep day and 2 rest days.

$$\prod_{j}^{j+4} w_{i,3,j} = 0, \ \forall\, i = \{1, \ldots, n\}, \forall\, j = \{1, \ldots, D - 5 + 1\} \quad (2c)$$

$$\prod_{j}^{j+5} w_{i,3,j} = 0, \ \forall\, i = \{1, \ldots, n\}, \forall\, j = \{1, \ldots, D - 5 + 1\} \quad (2d)$$

Where,
$w_{i,3,j}$ = 1 if nurse i is working on night shift on day j, and 0 otherwise

**H3**: Nurse Requirement Constraint

The number of nurses must be at least $R_{s,t,j}$ for a particular shift s in a ward t on day j

$$\sum_i (h_{i,t} * w_{i,s,j}) \geq R_{s,t,j}, \ \forall\, s = \{1,2,3\}, \forall j = \{1, \ldots, D\} \quad (3)$$

Where,
$R_{s,t,j}$ = minimum number of nurses required on shift s in ward t on day j
$h_{i,t}$ = 1 if nurse i is trained for ward t, 0 otherwise
$w_{i,s,j}$ = 1 if nurse i is working on shift s on day j, and 0 otherwise

**H4**: Annual Leave and Training Leave Constraint

For nurse i who has applied for leave on day j, we let $L_{i,j} = 1$

$$\sum_s w_{i,s,j} = (L_{i,j})', \ \forall\, i = \{1, \ldots, n\}, \forall j = \{1, \ldots, D\} \quad (4)$$

Where,
$w_{i,s,j}$ = 1 if nurse i is working on shift s on day j, and 0 otherwise
$L_{i,j}$ = 1 if nurse i has applied for leave on day j, and 0 otherwise
$L_{i,j}'$ = complement of $L_{i,j}$

**H5**: No Consecutive Shift Constraint

Consecutive shifts including Morning-Afternoon, Afternoon-Night, and Night-Morning are not permitted in the schedule.

$$w_{i,1,j} \cdot w_{i,2,j} = 0, \ \forall \ i = \{1, \dots, n\}, \forall j = \{1, \dots, D\} \quad (5a)$$

$$w_{i,2,j} \cdot w_{i,3,j} = 0, \ \forall \ i = \{1, \dots, n\}, \forall j = \{1, \dots, D\} \quad (5b)$$

$$w_{i,3,j} \cdot w_{i,1,j+1} = 0, \ \forall \ i = \{1, \dots, n\}, \forall j = \{1, \dots, D-1\} \quad (5c)$$

The constraints set out are established mainly by regulatory requirements which do not permit for any adjustments. Even though **H4** can be adjusted or controlled by the planning staffs, the labor laws prohibit any actions that prevents an employee from consuming the entitle leave of absence.

Soft constraints are features of the scheduling of the nursing staffs. While they are more flexible than the hard constraints, they are valuable in assisting the planners to cater to the preferences of the nursing staffs that will be more acceptable and less prone to changes. The soft constraints are set out as below.

**S1**: An afternoon shift should not be followed by a morning shift the next day.

**S2**: Continuous rest days are not preferred unless it is meant to follow 4 consecutive night shifts.

**S3**: A rest day should be followed by more than 1 work day.

**S4**: Three or four consecutive night shifts are not preferred.

These soft constraints are established with advice from the planning staffs. Most of the soft constraints are adjustable and new soft constraints can be defined and added, and it is up to the planning staffs' preferences to decide on the choice of action. For example, S4 constraint might be otherwise acceptable to a nurse who prefers night shift over day shifts. Thus S4 constraint might result in a positive preference value as opposed to a penalty cost in certain situation. S2 might be preferred if the nurse wishes to take a small holiday which is again a preference issue.

To capture the nurses' preferences, there is a list of preferences or soft constraints issue that can be drawn and sent to the nurses for their inputs. Below is a sample of the inputs, where a higher penalty cost will indicate low preference, and vice versa. We have assumed that all nurses have the same assignment of penalty cost. However, we are confident that if the nurses assignment of penalty cost are different, a better solution is almost guaranteed, as such complementary preferences will offer better combinations of schedules which can satisfy most nurses.

| Shift combination | Penalty Cost |
|---|---|
| N-N-PM | 500 |
| N-N-RD | 50 |
| N-PM | 25 |
| N-RD | 25 |
| PM-AM | 10 |
| RD-RD | 10 |
| N-N-N | 5000 |

Table 1: Example of Preference Cost Input

## 4.0 Greedy Double Swap Heuristic

The Greedy Double Swap Heuristic (GDSH) is a simple heuristic that can be viewed as a variant of the Greedy Randomized Adaptive Search Procedures (GRASP) (Resende and Pitsoulis, 2002). In GRASP, the algorithm creates a sub-space of feasible solution before creating an additional search localized for local minima. The process is repeated many times to look for the most optimum solution.

The GDSH has 2 phases. The GDSH differs from GRASP in that it does not create an initial list of potential solution in several subspaces but instead begin by searching for one single possible subspace of solution. Once the subspace of solution is found, the GDSH will perform the localized search like GRASP. Another difference between the two algorithms is that the GRASP searches for the best solution within the subspace of solutions by changing to a new solution which might be totally different from the previous one. GDSH swaps two variables (e.g. two shifts for the same nurse) at a single time to search for a better solution, which could be in another subspace.

One of the key challenges of heuristic is the tendency to get lock in a small sub-space of solution with local minima due to the nature of single variable change algorithms. This poses a particular problem in the NSP. Typically, once the algorithms move into local minima, unless there is a drastic change in the solution, the solution will remain as the local minima. However, this is less likely in the case of GDSH. The two variables swap allows the solution to move out from one subspace of solution to another subspace. This creates more opportunity for the algorithm to converge on the global minima. The GDSH also builds into it a layer of human logic by initiating a preliminary swapping algorithm, swapping two shifts for the same nurse, to reach the solution subspace before moving into Phase 2 where second swap which imitates the human behavior in changing the shifts between two nurses to meet the constraints. This swapping process is enhanced by the identification of the component that is the least contributing to the penalty cost objective function in an attempt to reduce the total penalty cost.

There are a few stages to the scheduling process. The Pre-Scheduling stage will calculate whether the number of nurses is sufficient to cover the duty requirements of the period, which is total supply satisfies total demand. This is critical as insufficient supply of nurses will violate the hard constraints which are regulatory in nature. Once the Pre-Scheduling component has determined that supply meets or exceeds demand, Phase 1 of the algorithm will begin where the initial swapping of shifts for the same nurse occurs and the algorithm attempts to look for a subspace of solution that meets the hard constraints. Once the hard constraints are met, the GDSH moves into Phase 2.

At Phase 2, the algorithm focuses on swapping the shifts between two nurses to minimize the impact of any soft constraints while enforcing the hard constraints.

The impact is quantified as a penalty cost added to the penalty cost objective function which needs to be minimized. The optimization phase will continue to iterate until convergence is reached or when no better solutions can be found after 100,000 iterations. The GDSH is described mathematically using the pseudo-code as shown below.

Greedy Double Swap Algorithm

*Initialization*

1. Read in the parameters.
2. Calculate the manpower supply and demand.
    i.   If demand exceeds supply, terminate.
    ii.  Else, continue.
3. Initialize an initial schedule of the nurse based on the nurses required number of working days. This initial solution may violate the hard constraints.
4. Input the required days for leave of absence.

*Phase 1 - First swap between shifts for the same nurse*

1. Randomly select a nurse n.
2. Swap the shift for day i and day j of nurse n.
3. Check for hard constraints condition.
    i.   If the percentage of hard constraints met improves, then use the new solution.
    ii.  Else, use back the old solution.
4. Repeat steps 1 to 3 until 100% of hard constraints are met or when iterations exceed 1,000,000.

*Phase 2 - Double Swap with Minimization of Penalty Cost*

1. Calculate the total penalty cost incurred by every nurse.
2. Search for the nurse n which contributes most to the penalty cost objective function.
3. Select another nurse m randomly.
4. Randomly select two days i and j.
5. Swap the shifts of nurse n and m for days i and j.
6. Check for hard constraints conditions.
    i.   If there is violation, reverse the move.
    ii.  Else, continue.
7. If 6 ii. is executed, check for penalty cost objective function improvement.
    1. If there is no improvement, reverse the move.
    2. Else, continue.
8. Repeat steps 1 to 8 until penalty cost objective function = 0 or iterations exceed 10,000.

The above algorithm can be coded into any spreadsheet tools using VBA, Python or OpenOffice Basic. We have implemented the algorithm in an open source package called LibreOffice which is currently supported by several releases of Ubuntu and its variants. With the heuristic defined, we will proceed to conduct a Monte Carlo simulation to determine its efficacy and efficiency.

## 5.0 Experiments

To examine the efficacy and efficiency of the algorithm, we will conduct two experiments. In the first experiment, we will compare the performance of GDSH with two NLP algorithms using a simplified version of the NSP problem. The simplified problem will involve only the hard constraints which can be modeled using mathematical programming easily. We will compare if each method is able to solve the NSP problem, and how fast it takes. Once we have determined the efficacy and efficiency of each approach, we will run simulations of the actual problem and attempt to measure their performance.

In the simplified experiment, we used a case of 6 nurses without any specialist training for a single ward for a 5-day schedule. There are only hard constraints **H3** as detailed in the section above and the objective function is to ensure that all the constraints are met. We performed 100 simulation runs for each solution method and compute the average time taken to solve the problem, as well as how far the feasibility of the solution is satisfied. Feasibility here is defined as the percentage of the constraints met.

| Solution Method | Time(Seconds) | Feasibility (%) |
|---|---|---|
| SCO Evolutionary NLP | 1765 | 90.29 |
| DEPS Evolutionary NLP | 1748 | 87.27 |
| GDSH | 368 | 100 |

Table 1: Simplified Problem Monte Carlo Results

From the results, we can observe that both NLP solution methods did not manage to solve the problem, even after increasing the number of iterations. Given the inability of both NLP solution methods to solve the simplified problem, it is clear that both methods will not be able to solve the actual problem within an acceptable time frame.

In the actual experiment, we used the case of 15 nurses with specialist requirements for a single ward for a 14-day schedule. There are various requirements for the number of nurses of a particular specialized skill to be present. All the hard and soft constraints defined in the earlier section are implemented. The GDSH will first ensure that all the hard constraints are met in Phase 1 before attempting to minimize the cost Phase 2. Various soft constraints resulting in undesirable schedule structures are assigned cost which will be calculated for individual nurse's schedule as well as the entire ward's schedule. We will run 100 simulation run of the GDSH to examine the time taken as well as the feasibility of the solution.

| Stages | Time (Seconds) |
|---|---|
| Phase 1 | 60 |
| Phase 2 | 318 |

| **Objective Value** | 1257 |
|---|---|

Table 2: Full Problem Monte Carlo Results

From the table, we can observe that the algorithm generates results very quickly and most of the results are able to obtain good feasible schedules. There were instances where the objective function is not minimized sufficiently, but with the speed at which the algorithm can run, it allows multiple re-runs to obtain a better solution. However, we note that 10% of the simulation runs resulted in failure or objective function value which are extremely large.

## 6.0 Conclusions

The GDSH serves as an alternative solution approach to the NSP problem. The proposed algorithm has the ability to create a possible schedule in a very short time, usually within minutes. Such a fast solution method can be implemented in practice and allows for quick reaction to any schedule changes. The GDSH also searches for a sub-space of solution before optimization to speed up the process of obtaining a solution. By allowing individual nurses to assign their preferences of the soft constraints using penalty costs, the algorithm attempts to satisfy most preferences by minimizing the penalty cost. In addition, the flexible nature of the heuristic also provides easy assimilation of soft constraints into the problem.

By using the small experiment, we have shown that GDSH can solve the NSP problem better than mathematical programming. The heuristics has also provides solutions which are fairly good and within a reasonably short time.

There is still room for improvement. One of the problems with the swapping algorithm is that it cannot account for cases where multiple swaps among several variables are needed to meet the hard constraints. Investigations into the failure cases indicate that the algorithm failed due to a need for swapping the shifts on different days for more than one nurse in Phase 1 in order to move out of the infeasible solution subspace. As such swap moves are not easily identifiable, a more in-depth modification to the algorithm is needed to address this problem. At the same time, the preferences of the nurses are not always the same. In this paper, we have assumed a preference structure which is uniform among the nurses, and additional experiments need to be done on cases where the nurses preferences are different.

We conclude that the GDSH is a simple and easily implementable algorithm in open source spreadsheet software. Although the current problem is a NSP, the algorithm can be extended to any scheduling problem.

# References

Aickelin, U., & Dowsland, K. (2004). An indirect genetic algorithm for a nurse-scheduling problem. Comput. Oper. Res., 31 (5), 761–778.

Azaiez, Al-Sharif, (2005). A 0-1 goal programming model for nurse scheduling. Computers & Operations Research - CoR , vol. 32, no. 3, pp. 491-507.

Brusco, M., & Jacobs, L. (1993). A simulated annealing approach to the solution of flexible labour scheduling problems. The Journal of the Operational Research Society, 44 (12), 1191–1200.

Burke, E., Causmaecker, P., Berghe, G., & Landeghem, H. (2004). The state of the art of nurse rostering. Journal of Scheduling, 7 , 441–499.

Chen, J.-G., & Yeung, T. (1993). Hybrid expert-system approach to nurse scheduling. Computers in Nursing, 11 , 183–192.

Coello-Coello, C. (2002). Theoretical and numerical constraint-handling techniques used with evolutionary algorithms: a survey of the state of the art. Computer Methods in Applied Mechanics and Engineering, 191 (11-12), 1245 – 1287.

Dowsland, K. (1998, April). Nurse scheduling with tabu search and strategic oscillation. European Journal of Operational Research, 106 (2-3), 393–407.

Dowsland, K., & Thompson, J. (2000, Jul). Solving a nurse scheduling problem with knapsacks, networks and tabu search. The Journal of the Operational Research Society, 51 (7), 825–833.

Li, J., & Aickelin, U. (2003). A bayesian optimization algorithm for the nurse scheduling problem.

Okada, M., & Okada, M. (1988). Prolog-based system for nursing staff scheduling implemented on a personal computer. Comput. Biomed. Res., 21 (1), 53–63.

Osogami, T., & Imai, H. (2000). Classification of various neighborhood operations for the nurse scheduling problem. In ISAAC '00: Proceedings of the 11th International Conference on Algorithms and Computation (pp. 72–83). London, UK: Springer-Verlag.

Pelikan, M., Goldberg, D. E., & Cant´u-Paz, E. (1999). BOA: The Bayesian optimization algorithm (Technical Report IlliGAL 99003). Urbana, IL: Illinois Genetic Algorithms Laboratory, University of Illinois at Urbana-Champaign.

Tien, J., & Kamiyama, A. (1982). On manpower scheduling algorithms. SIAM Review, 24 (3), 275–287.

Ying-Shiuan You, Tian-Li Yu, and Ta-Chun Lien. (2010). Psychological Preference-based Optimization Framework: An Evolutionary Computation Approach for Constrained Problems Involving Human Preference . Master Thesis, TEIL Working paper.